\documentclass{article}
\usepackage{emulateapj}
\usepackage{times}

\usepackage{psfig}

\newcommand\msun{{\,M_\odot}}

\newcommand\fe{Fe K$\alpha$~}

\newcommand\fx{F_{\rm x}}

\newcommand\fdisk{F_{\rm d}}

\newcommand\ptot{P_{\rm tot}}

\newcommand\teff{T_{\rm eff}}

\newcommand\fmag{F_{\rm mag}}
\newcommand\pmag{P_{\rm mag}}
\def\>{$>$}
\def\<{$<$}

\def\simlt{\lower.5ex\hbox{$\; \buildrel < \over \sim \;$}}
\def\simgt{\lower.5ex\hbox{$\; \buildrel > \over \sim \;$}}
\def\sqr#1#2{{\vcenter{\hrule height.#2pt
      \hbox{\vrule width.#2pt height#1pt \kern#1pt
         \vrule width.#2pt}
      \hrule height.#2pt}}}

\begin{document}


\title{On the ``failure'' of the standard model for Soft X-ray Transients
in quiescence.}

\author{Sergei Nayakshin\altaffilmark{1} and Roland
Svensson\altaffilmark{2}}
\altaffiltext{1}{USRA and NASA/GSFC, LHEA, Code 661, Greenbelt, MD,
20771 (serg@milkyway.gsfc.nasa.gov)}
\altaffiltext{2}{Stockholm Observatory, SE-133 36 Saltsj\"obaden, Sweden
(svensson@astro.su.se)}


\begin{abstract}
It is currently believed that the ``standard'' accretion disk theory
under-predicts the observed X-ray luminosity from Soft X-ray
Transients (SXT) in quiescence by as much as 4 to 6 orders of
magnitude. This failure of the standard model is
considered to be an important argument for the existence of
the alternative mode of accretion -- Advection Dominated Accretion
Flows (ADAF) in astrophysics, since these flows allow a much higher
level of X-ray emission in quiescence, in agreement with the
observations. Here we point out that, in stark contrast to
steady-state standard disks, such disks in quiescence (being
non-steady) produce most of the X-ray emission very far from the last
stable orbit.  Taking this into account, these disks {\em can}
accommodate the observed X-ray luminosities of SXTs rather naturally.
Our theory predicts that \fe lines from standard accretion disks in
quiescence should be narrow even though the cold disk goes all the way
down to the last stable orbit.
\end{abstract}


\section{Introduction}\label{sect:intro}

Soft X-ray transients are mass transfer binaries which periodically
undergo outbursts in which their luminosity increase by several orders
of magnitude (e.g., see reviews by Tanaka \& Lewin 1995; van Paradijs
\& McClintok 1995; Wheeler 1996; Cannizzo 1998).  The compact object
may be either a neutron star (NS) or a black hole (BH). SXTs are
particularly interesting given the ongoing debate about the structure
of the inner accretion disk. In quiescence, the luminosity of SXTs
seems to be much higher (e.g., Verbunt 1995; Lasota 1996, Yi et
al. 1996; Lasota, Narayan \& Yi 1996a; Narayan, McClintock \& Yi 1996)
than allowed by the standard model (Shakura \& Sunyaev 1973; expected
to be modified for SXTs due to hydrogen-ionization instability --
e.g., Mineshige \& Wheeler 1989 [MW89]). Lasota, Narayan \& co-workers argued
that because the ``standard'' theory predicts that the inner accretion
disk is essentially empty of mass in quiescence and accretes at a very
small accretion rate, $\dot{M}_{\rm in}$, it cannot reproduce the
observed X-ray luminosity $L_{\rm x, obs}$: typical numbers are $0.1
\dot{M}_{\rm in} c^2\sim (10^{-4} - 10^{-6})\,\times L_{\rm x,
obs}$. They conclude that SXTs in quiescence clearly invalidate the
standard model.

However, in this Letter, we argue that this argument is applicable
only to ``bare'' standard accretion disks -- those without an
overlying hot corona.  We find that quiescent standard disks {\em with
a corona} should produce most of their X-ray emission at very large
radii precisely due to the fact that the inner disk is almost empty of
mass. At large radii, the standard theory of hydrogen-ionization
instability allows much larger accretion rates, and the observed X-ray
luminosities can be easily reproduced if the corona reprocesses $\sim
10$\% of the disk energy budget. We also make a testable prediction --
\fe lines from quiescent accretion disks should be narrow (even though
these disks extend down to the last stable orbit or the surface of the
compact star).

\section{Hydrogen ionization instability: basic facts}
\label{sect:inst}

Because the argument of Lasota (1996) against the standard model in
SXTs is qualitative, we will address it with a qualitative model as
well.  The hydrogen ionization instability of accretion disks has been
studied for more than two decades for a range of central objects --
White Dwarfs (WD), NS, and galactic and extra galactic BHs (e.g.,
Hoshi 1979; Smak 1982; 1984; MW89; Mineshige \& Shields 1990 [MS90];
Cannizzo 1993; Siemiginowska, Czerny \& Kostyunin 1996
[SCK96]). Cannizzo (1998, \S 2 \& 3) gives a short overview of the
development of the theory and an easily accessible and enlightening
discussion of the background physics.  Here we will only sketch the
basic principles of the instability to the extent needed to explain
our main point.  The behavior of the unstable disk depends on the
so-called ``S-curve'' that relates the local disk effective
temperature and the column density (see Figures in Smak 1982, MS90,
SCK96; and Fig. 31 in Frank, King \& Raine 1992). The curves are
different for different radii, but luckily only slightly. The steady
state Shakura-Sunyaev solution predicts that the disk effective
temperature, $\teff(R)$, changes as $R^{-3/4}$ with radius. In
general, this predicts that for small $R$, hydrogen is completely
ionized, then there should be a range of radii where hydrogen is
partially ionized, and finally, at large $R$, hydrogen should
recombine completely.

However, it turns out that the partially ionized solution is thermally
unstable. In terms of the S-curve, this region is between points C and
E (see Fig. 31 in Frank et al. 1992). Numerical simulations show that
the unstable disks oscillate between two quasi-stable states. The
accretion rate is not constant with time and radius, but the
time-averaged accretion rate equals that at which the matter is
supplied at the outer edge of the disk, $\dot{M}_0$. One of these
states -- the region between points E and D on the S-curve -- is the
outburst state in which hydrogen is completely ionized. The other
thermally stable state -- the one between points B and C on the
S-curve in Fig. 31 of Frank et al. 1992 -- is the quiescent state when
hydrogen is neutral. To remain in this state, the mid-plane gas
temperature should be kept below $\sim 10^4$ Kelvin, which in turn
requires $\teff(R) < T_C(R) \simlt {\rm few} \times 10^3 $ (depending
on the black hole mass, $M$, and viscosity prescription).

Time-dependent (full disk) calculations of many authors show that when
the unstable region of the disk returns to quiescence, it settles
around point B on the S-curve, which is a little below the maximum
quiescent temperature and surface density (point C). The accretion
rate corresponding to point B, $\dot{M}_B(R)$, is lower than
$\dot{M}_0$, so that the matter starts accumulating, most quickly at
the outermost part of the unstable region. At a future time, this
piling up of the mass will lead to $\dot{M}(R)> \dot{M}_C$ at some
radius (equivalently, $\teff(R)> T_C(R)$), and the outburst will be
triggered at that radius (see Cannizzo 1998, \S 3).

Fortunately, for the present analysis, these details are mostly
irrelevant.  What is important for us here is that both local disk
calculations and full disk time-dependent ones show that the gas effective
temperature in quiescence is between $T_B(R)$ and $T_C(R)$; that
$T_B(R)$ is typically smaller than $T_C(R)$ by a factor of 2 or less,
and that both of these functions are extremely weakly dependent on
$R$. In particular, the scaling law given by Smak (1984), MS90 and
SCK96 is $T_C(R)\propto R^{-q}$, where $q$ is between 0.08 and 0.12.
Full disk calculations give temperature profiles slowly increasing and
then decreasing with $R$ (e.g., Fig. 13 and 16 in MS90; Fig. 10 in
SCK96), which is probably due to the fact that the outer disk fills up
quicker and hence lies closer to point C than the inner disk
does. There are also observations of the disk effective temperature in
quiescence for eclipsing dwarf novae, systems to which the
hydrogen-ionization instability was applied first. These observations
show that $\teff(R)$ is either a constant or a slowly decreasing
function of $R$ (Fig. 11 in Wood et al. 1986; Fig. 14 in Wood et
al. 1989; and Figs. 5 \& 6 in Bobinger et al. 1997).

While the exact run of $\teff(R)$ with radius depends on details of
the model and the object under consideration, {\em all} of the above
observational and theoretical papers show that $\teff(R)\sim $ const
in quiescence within a factor no larger than $2$. This is quite
remarkable given that the radii of the unstable region can span some
two or more orders of magnitude: the steady-state Shakura-Sunyaev
solution with $\teff(R)\propto R^{-3/4}$ would predict a change in
$\teff$ by about a factor of $30$. Hence, below we will assume that
$\teff(R)\sim T_U \equiv$ const $\simlt T_C\sim $ few $\times 10^3$ K
in quiescence. Physically, the near isothermality of the disk in
quiescence is due to the fact that the ionization state of hydrogen is
a very sensitive function of temperature, and large changes in radius
cause only a slight difference in $T$. For example, the opacity is an
extremely strong function of midplane disk temperature $T_d$: in the
temperature range $6000 \simlt T_d \simlt 8000$ K, the opacity
$\propto T_d^b$ where $b$ is as large as 10--20 (see Cannizzo \&
Wheeler 1984).

Of course, the effective temperature is approximately constant only
inside the unstable disk region. Let us call the corresponding radius
$R_U$.  For larger radii, the disk is always non-ionized and so it
follows the steady state Shakura-Sunyaev solution, i.e.,
$\teff(R)\propto R^{-3/4}$.  Summarizing,
\begin{equation}
\teff(R) \simeq \cases{ T_U, & if\, $R \, < R_U$ \cr T_{\rm SS}(R) &
otherwise \, ,\cr}
\label{tu}
\end{equation}
where $T_{\rm SS}(R)$ is the Shakura-Sunyaev effective temperature of
the disk corresponding to the given radius, accretion rate $\dot{M}_0$
and the black hole mass. The thermal disk flux is simply given by
$\fdisk(R) = \sigma_B \teff^4(R)$.

What are the implications of this for the accretion rate through the
disk? At $R\gg R_s\equiv 2GM/c^2$, $\fdisk(R) = (3/8\pi)
GM\dot{M}(R)/R^3$, which can be solved for $\dot{M}(R)$:
\begin{equation}
\dot{M}(R) \simeq \cases{ \dot{M}_0 (R/R_U)^3, & for $R < R_U$ \cr
\dot{M}_0 & for $R > R_U$\cr}
\label{mofr}
\end{equation}
That is, the accretion rate scales as $\dot{M}(R)\propto R^3$ within
the unstable radius. Physically, this comes about from the fact that
it is most difficult to keep the hydrogen recombined for smaller radii
because the gravitational and viscous energy release scale as
$R^{-3}$, and hence the disk has to be most strongly depleted for the
smallest $R$ to remain quiescent (see also Cannizzo 1998).

\section{X-rays do not always come from the inner disk}\label{sect:cr} 

A frequent implicit or explicit assumption in the common logic about the
X-ray emitting region(s) is the following. Since X-rays are the
``hottest'' part of the overall disk spectrum, and since the Shakura \&
Sunyaev (1973) model predicts that the disk is hottest in its
innermost region, then that has to be the region where the X-rays are
produced. Seemingly innocent, this suggestion is actually appropriate
for the ADAF-type models (i.e., the inner-ADAF plus outer cold disk
models), but is actually entirely foreign for standard disks with
coronae in quiescence.

To show this, let us first assume without any justification that the
corona above the disk exists at all radii, and that the fraction of
power transferred into the corona, $f$, is independent of radius
(e.g., Svensson \& Zdziarski 1994). That is, $\fx(R) = f \fdisk(R)$.
Next, we define the X-ray luminosity per decade of radius, as
$L_x(R)\equiv R^2 \fx(R)$, which obviously scales according to
\begin{equation}
L_x(R) = \cases{ f L_U (R/R_U)^2, & for \,  $R < R_U$ \cr
f L_U (R/R_U)^{-1} & for $R > R_U$ \cr}\, ,
\label{lofr}
\end{equation}
where $L_U = \fdisk(R_U) R_U^2$.  Therefore, {\em most} of the viscous
energy liberation {\em and X-ray production} for a quiescent disk
occurs at $R\sim R_U \gg 3 R_S$, if $f=$ const. The inner disk in the
standard model is simply very dim and for many purposes may be
unobservable {\em at any wavelength} in quiescence.

Is it appropriate to assume $f\simeq $ const? To date, detailed MHD
simulations of accretion disks have not yet addressed this question,
because global MHD disk simulations are forbiddenly numerically
extensive, and hence calculations are limited to a small region in
radius (see, e.g., Miller and Stone 2000). In the absence of these, we
will make a simple analytical argument. If one defines the
$\alpha$-parameter through $\nu = \alpha c_s H$, where $c_s$ and $H$
are the sound speed and the pressure scale height, respectively, then
the total energy generated by the disk per unit surface area is $2
\fdisk(R) = (9/8)\, (GM/R^3)\, \nu \Sigma = (9/8)\, \alpha c_s \ptot$,
where $\Sigma$ is the vertical mass column density, and $\ptot$ is the 
total disk mid-plane pressure. The flux of magnetic energy, $\fmag$,
delivered by buoyancy to the disk surface and then presumably
transformed into X-rays by reconnection is $ \fmag \sim v_b
\langle\pmag \rangle$, where $v_b\simlt c_s$ is the average buoyant
velocity, and $\langle\pmag\rangle$ is the average magnetic field
pressure ($\sim$ energy density). The viscosity parameter $\alpha$ was
defined by Shakura \& Sunyaev (1973) to be $\alpha \simeq \alpha_t +
{\langle{P_{\rm mag}}\rangle\over P_{\rm tot}}$, where the first term,
$\alpha_t$ is the turbulent viscosity coefficient. It is now believed
that the magnetic viscosity is significantly more important than the
turbulent one (e.g., Hawley, Gammie, \& Balbus 1995; 1996), and that
$\alpha \sim \langle{P_{\rm mag}}\rangle/ P_{\rm tot}$, and hence
\begin{equation}
\fmag \sim v_b \alpha \ptot \simlt \alpha c_s \ptot \sim \fdisk\; .
\label{fm2}
\end{equation}
In other words, this equation shows that $f$ is indeed expected to
be a constant with respect to $R$.

Further, if corona is powered by magnetic flares (e.g., Galeev, Rosner
\& Vaiana 1979; Haardt, Maraschi \& Ghisselini 1994; Beloborodov 1999;
Miller \& Stone 2000), then it is clear that the resulting X-ray
spectra have little to do with the temperature of the underlying
disk. The spectra depend primarily on the Thomson depth of the hot
plasma and {\em its} temperature. The temperature within a flare is
controlled by the ratio of the X-ray flux to the disk flux,
$\fx/\fdisk$, (e.g., Stern et al. 1995; 
Poutanen \& Svensson 1996),
plus the exact (unknown) geometry of the reconnection site. Therefore,
it is entirely possible that a magnetic flare that occurs at $R\sim
1000\, R_S$ will have an X-ray spectrum similar to, or even harder
than that of a flare at $R \sim 10 R_S$, especially for the quiescent
disks.

To summarize this discussion, there is nothing special about the
innermost region of the disk for quiescent accretion disks. If there
is a corona above the disk, then most of the emission at any
wavelength will be produced at $R\sim R_U$ [Note that in a steady
Shakura-Sunyaev disk, most of the viscous energy release happens in
the inner disk, and this is why it is safe to assume that X-rays come
from very small radii {\em for such disks}]. Since there are several
parameters that will determine the resulting X-ray spectra (e.g.,
reconnection rate, height and geometry of the X-ray source, Thomson
depth of the X-ray emitting plasma, etc.), it is difficult to
calculate the spectra model-independently. Nevertheless, we note that
variations of these parameters are sure to yield a broad range of
spectral indices (see, e.g., Stern et al. 1995; Poutanen \&
Svensson 1996), and hence it may be entirely possible to reproduce the
observed spectra of SXTs.

\section{X-ray ``Problem'' for standard disks in
quiescent SXTs}\label{sect:sxts}

Currently, it is believed that there exists a clear-cut ``evidence''
for a major failure of the standard disk instability model for the
quiescent state accretion disks around dwarf novae, neutron stars and
galactic black holes (SXTs). The popularity of the ADAF model itself
in part rests on this fact because it is claimed to be the only model
that can explain self-consistently the level of X-rays in quiescence
for the systems mentioned above. Let us for a moment accept the
suggestion that X-rays do emanate from the innermost region and follow
the arguments of Lasota (1996), as well as many of other authors (e.g.,
Lasota et al. 1996a; Yi et al. 1996). The requirement for the
disk material to be below the instability point C on the S-curve
limits the accretion rate to be (Lasota 1996):
\begin{equation}
\dot{M}_{\rm m}(R) \simeq 2.76 \times 10^3 t_9^{-1}\, r_7^{3.11}
M_1^{-0.37}\, \alpha^{-0.79}\; {\rm g\ s}^{-1}\;,
\label{lasota1}
\end{equation}
where $M_1$ is the mass of the black hole in units of $10 \msun$,
$t_9$ is the recurrence time in units of $10^9$ s, and $r_7 \equiv
R/10^7$~cm. Numerical models (e.g., MW89) satisfy this
requirement. This condition (eq. 5) is similar to the simpler one
given by equation (\ref{mofr}) with a scaling $\dot{M}_{\rm
m}(R)\propto R^{3.11}$ instead of $R^3$.  Using a $10$~\% efficiency
for conversion of mass accretion rate into X-rays, Lasota (1996) noted
that two SXTs, A0620-00 and V404~Cyg, have been detected at the
accretion rates $\sim 10^{10}$ and $\sim 3\times 10^{12}$ g s$^{-1}$
(see Verbunt 1995). This estimate is 4-6 orders of magnitude larger
than that given by equation (\ref{lasota1}).

However, as we argued in \S \ref{sect:cr}, most of X-ray emission
comes from $R\sim R_U \gg 3 R_S$ for standard disks in
quiescence. Approximately, we have
\begin{equation}
L_x \simlt f\, \frac{GM\dot{M}(R_U)}{R_U} = f \frac{R_S}{2 R_U}
\dot{M}(R_U)\, c^2\;,
\label{lasota2}
\end{equation}
where $\dot{M}(R_U)\sim \dot{M}_0$.  Since in quiescence
$\dot{M}(R_U)\simlt \dot{M}_{\rm m}(R_U)$, we estimate
\begin{equation}
L_x \simlt 3.7 \times 10^{23} \, f\, t_9^{-1}\, r_7^{2.11}
M_1^{0.63}\, \alpha^{-0.79}\; {\rm erg\ s}^{-1}\;.
\label{lasota3}
\end{equation}
Clearly, by allowing $r_7 = R_U/10^7 {\rm cm}$ to be much greater
than unity, we can obtain much higher $L_x$ than the estimate based on
10\% efficiency and $r\sim $few.

Let us now use theoretical and observational constraints on the
parameters in equation (\ref{lasota3}) to determine $L_x$ more
accurately for  A0620-00. From observations, we
know that $L_x\sim 0.1 L_{\rm bb}$, where $L_{\rm bb}\simeq 10^{32}
{\rm erg\ s}^{-1}$ is the blackbody optical luminosity of the disk as
inferred by McClintock et al. (1995). This means that we need to
choose $f\simeq 0.1$. Further, analyses of H$\alpha$ and $H\beta$
spectroscopic data (Marsh et al. 1994) indicate that the radius where
the Balmer line luminosity peaks is $\sim 5\times 10^{10}$.  This
should roughly equal $R_U$ since that is where most of the luminosity
is produced in our model. Therefore,
\begin{equation}
L_x \sim 9.4 \times 10^{30} \left(\frac{f}{0.1\, t_9}\right)\,
\left(\frac{M}{5\msun}\right)^{0.63}\,
\alpha_1^{-0.79}\; \frac{\rm erg}{\rm s}\;,
\label{me}
\end{equation}
where $\alpha_1\equiv \alpha/0.1$.  This value of $L_x$ is in a very
good agreement with the observed $L_x\sim 10^{31} {\rm erg\ s}^{-1}$.
A further independent check of the self-consistency of this picture can be
made by computing the temperature $T_U$, from $\sigma_B T_U^4 = (1-f)
(3/8\pi) GM\dot{M}_0/R_U^3$. For $\dot{M}_0 \simeq 10^{16}$ g s$^{-1}$
(see McClintock et al. 1995), we find
\begin{equation}
T_U = 3.2 \times 10^3\, \left(\frac{M}{5\msun}\right)^{1/4}\;.
\label{tucheck}
\end{equation}
This temperature estimate agrees quite well with the expected
temperature at the lower kink of the S-curve (see Figs. in Smak 1982,
MS90, SCK96).

\section{Discussion}\label{sect:conclusion}

In this paper we re-examined the argument given by Lasota (1996),
Lasota et al. (1996a) and Yi et al. (1996) with which they
rule out the standard accretion disks extending down to the last stable
orbit for quiescent states of SXTs. The argument is based on the assumption
that all X-rays are emitted very close to the last stable orbit. Because
the accretion rate through this region is so low in quiescence, the
standard theory appears to under-predict the observed X-ray
luminosities by orders of magnitude. However, we have shown that this
argument is only applicable to bare standard accretion disks, i.e.,
those that have no hot X-ray producing corona. Such disks are ruled
out already due to the fact that they do not produce hard X-rays at
all, and hence there is no reason to invoke them as a model for SXTs
in the first place (and not only for SXTs but for all X-ray emitting
accreting sources).  We have shown that if a corona exists at all
radii, then most of X-ray emission will come from large radii in
quiescence. At these radii, the maximum allowed accretion rate -- the
maximum one at which hydrogen is neutral -- is much
larger. Accordingly, much larger X-ray luminosities may be sustained
by such ``non-bare'' disks, in qualitative agreement with the X-ray
observations of quiescent SXTs.

Further progress in testing this and the ADAF models for SXTs can only
be made by a detailed comparison of theoretical and observed
spectra. While there is a large body of literature on the spectra of
the ADAF model for SXTs (e.g., see the recent review by Lasota 2001)
and none for the model proposed here, both of these face serious
theoretical challenges and hence neither model can be preferred (in
our opinion). In particular, standard disks with magnetic flares is
perhaps the most viable model for {\em luminous} accreting black
holes, both AGN and GBHCs (e.g., Galeev et al. 1979; Beloborodov 1999;
Nayakshin \& Dove 2001; Done \& Nayakshin 2001).  However, it is not
clear at all how spectra of the high luminosity magnetic flares (e.g.,
Poutanen \& Svensson 1996) should be modified when applied to
SXTs. Ironically, the uncertainties in the theory may be great enough
to preclude both positive or negative statements about the
``standard'' model for SXTs, if these require detailed spectra.

The ADAF model, on the other hand, has fundamental difficulties in the
modeling of the cold-disk ADAF transition, which is therefore often
done in an ad-hoc way.  Another major uncertainty of the theory is the
accretion rate through the disk as a function of radius
($\dot{M}(R)$). The earlier papers assumed $\dot{M}(R)=$~const.
However, this would only be the case if the ADAF solution joined the
cold disk at $R_t \simgt R_U$, where $\dot{M}(R)$ is indeed constant
(see eq. \ref{mofr}). Yet $R_t$ is treated as a free parameter of the
model with $\dot{M}$ being independent of it (e.g., Lasota et
al. 1996b; Yi et al. 1996). From our analysis above, we
believe that if one chooses $R_t < R_U$, the accretion rate through the
ADAF region should be smaller by the factor $(R_t/R_U)^3$ (see also Wheeler
1996, on a similar point). Since the ADAF luminosity scales as $\dot{M}^2$
for small accretion rates, this factor translates into the sixth power
of $R_t/R_U$ in $L_x$. In addition, the issue is further complicated
by the more recent discovery of the importance of convection and
winds in the ADAF solution (e.g., Blandford \& Begelman 1999),
neglected in the earlier papers. Taken together, this implies that at
the present, the X-ray spectra of the ADAF model are no more well
determined than the spectra of magnetic flares.

One obvious consequence of our model is that because most of the X-ray
power is radiated away at $R\sim R_U$, \fe lines of quiescent SXTs
will have to be narrow even though the disk extends down to the last
stable orbit. While narrow-ish lines are also predicted by the ADAF
model, we believe that a detailed modeling and observations of these
lines are still the key to distinguish the two models. The ADAF model
produces the lines mostly due to collisional processes, and the line
is dominated by H- and He-like components (e.g., Narayan \& Raymond
1999) with the line centroid energy of $\sim$ 6.9 and 6.7 keV,
respectively. In the case of a cold disk plus corona model, depending
on the parameters of X-ray spectra we can expect the line centroid
energy to be either at $\sim 6.4$ or $6.7$ keV or be a combination of
these two (e.g., Nayakshin \& Kallman 2001). Highly ionized
collisionally excited \fe lines may also be expected based on the
analogy with \fe lines from solar flares (e.g., see Pike et
al. 1996). The ADAF model is however unlikely to produce a
neutral-like \fe line, because most of X-ray emission in the ADAF is
expected at small radii which yields little illuminating flux at large
radii where the cold disk is located.

Finally, we would like to stress the following general ``theorem''.
The absence of emission from the inner disk (whether this emission is
the \fe line emission, the X-ray continuum or the thermal disk
emission), is {\em not} an evidence for the absence of the standard
disk there. The particular case of SXTs discussed here proves this
point. Because hydrogen is neutral in quiescence, the inner disk is
expected to be very dim at all wavelengths and so it may be difficult
to discern in the data. This point is quite general as it may also be
relevant for other than hydrogen ionization disk instabilities. For
example, Nayakshin, Rappaport \& Melia (2000) have devised a specific
model for the radiation-pressure driven instabilities in the galactic
micro-quasar GRS~1915+105. In this model, the inner disk is also very
much dimmer during the ``quiescence'' than expected in the
steady-state Shakura-Sunyaev solution (see Fig. 13 in Nayakshin et
al. 2000). Therefore, one must not use the steady-state
Shakura-Sunyaev theory for unsteady disks -- e.g., low luminosity
systems, such as SXTs and Low Luminosity AGN (since one may suspect
that the reason these objects are dim is that they are in a quiescent
state), and those verifiably unsteady such as GRS~1915+105.

The authors acknowledge many useful discussions with John Cannizzo,
Chris Done and Demos Kazanas.

{}


\begin{thebibliography}{}

\bibitem[]{} Beloborodov, A. 1999, \apjl, 510, L123

\bibitem[]{} Blandford, R. D., \& Begelman, M. C. 1999, \mnras, 303, L1

\bibitem[]{} Bobinger, A., Horne, K., Mantel, K.-H., Wolf, S. 1997,
\aap, 327, 1023


\bibitem[]{} Cannizzo, J. K. 1993, in Accretion Disks in Compact
Stellar Systems, ed. J. C. Wheeler (Singapore: World Scientific), 6


\bibitem[]{} Cannizzo, J. K. 1998, in ASP Conf. Ser. 137: Wild Stars in 
The Old West, ed. S. Howel, E. Kuulkers, \& C. Woodward (San
Francisco: ASP), 308

\bibitem[]{} Cannizzo, J. K., \& Wheeler, J. C. 1984, \apjs, 55, 367

\bibitem[]{} Done, C., \& Nayakshin, S. 2001, \apj, 546, 419




\bibitem[]{} Frank, J., King, A., \& Raine, D. 1992, Accretion Power in
Astrophysics (Cambridge, UK: Cambridge University Press)

\bibitem[]{} Galeev, A. A., Rosner, R., \& Vaiana, G. S., 1979, \apj,
229, 318

\bibitem[]{} Haardt F., Maraschi, L., \& Ghisellini, G. 1994, \apjl,
432, L95

\bibitem[]{} Hawley, J. F., Gammie, C. F., \& Balbus, S. A. 1995, \apj,
440, 742

\bibitem[]{} Hawley, J. F., Gammie, C. F., \& Balbus, S. A. 1996, \apj,
464, 690

\bibitem[]{} Hoshi, R., 1979, Progr. Theor. Phys., 61, 1307



\bibitem[]{} Lasota, J.-P. 1996, in IAU Symp. 165: Compact Stars in
Binaries, ed. J. van Paradijs, E. P. J. van den Heuvel \&
E. Kuulkers (Dordrecht: Kluwer), 43

\bibitem[]{} Lasota, J.-P. 2001, New Astronomy Reviews, in press.

\bibitem[]{} Lasota, J.-P., Narayan, R., \& Yi, I. 1996a, \aap, 314, 813

\bibitem[]{} Lasota, J.-P., Abramowicz, M.A., Chen, X., Krolik, J.,
Narayan, R., \& Yi, I. 1996b, \apj, 462, 142


\bibitem[]{} Marsh, T. R., Robinson, E. L., \& Wood, J. H., 1994,
\mnras, 266, 137


\bibitem[]{} McClintock, J. E., Horne, K., \& Remillard, R. A. 1995,
\apj, 442, 358

\bibitem[]{} Miller, K. A., \& Stone, J. M. 2000, \apj, 534, 398

\bibitem[]{} Mineshige, S., \& Shields, G.A. 1990, \apj, 351, 47
(MS90)

\bibitem[]{} Mineshige, S., \& Wheeler, J.C. 1989, \apj, 343, 241 (MW89)



\bibitem[]{} Narayan, R., \& Raymond, J. 1999, \apjl, 515, L69

\bibitem[]{} Narayan, R., McClintock, J. E., \& Yi, I. 1996, \apj, 457,
821



\bibitem[]{} Nayakshin, S., \& Dove, J. 2001, \apj, in press

\bibitem[]{} Nayakshin, S., \& Kallman, T. 2001, \apj, 546, 406

\bibitem[]{} Nayakshin, S., Rappaport, S., \& Melia, F. 2000, \apj,
535, 798

\bibitem[]{} Pike, C.D. et al. 1996, ApJ, 464, 487

\bibitem[]{} Poutanen, J. \& Svensson, R. 1996, \apj, 470, 249



\bibitem[Shakura \& Sunyaev 1973] {sun73} Shakura, N. I., \& Sunyaev,
R. A. 1973, \aap, 24, 337

\bibitem[]{} Siemiginowska, A., Czerny, B., \& Kostyunin, V. 1996,
\apj, 458, 491 (SCK96)


\bibitem[]{} Smak, J. 1982, Acta Astr., 32, 199

\bibitem[]{} Smak, J. 1984, Acta Astr., 34, 161

\bibitem[]{} Stern, B. E., Poutanen, J.,
    Svensson, R., Sikora, M., \& Begelman, M. C. 1995, \apj, 449, L13

\bibitem[]{} Svensson, R. 1996, \aaps, 120C, 475


\bibitem[]{} Svensson, R. \& Zdziarski, A. A. 1994, \apj, 436, 599


\bibitem[]{} Tanaka, Y., \& Lewin, W. H. G. 1995, in X-Ray Binaries,
 ed. W. H. G.  Lewin, J. van Paradijs, \& E. P. J. van den Heuvel
(Cambridge, UK: Cambridge Univ. Press), 126



\bibitem[]{} van Paradijs, J., \& McClintock, J. E. 1995, in X-Ray
Binaries,  ed. J. van Paradijs, \& W. H. G.  Lewin, 
E. P. J. van den Heuvel (Cambridge, UK: Cambridge Univ. Press), 58


\bibitem[]{} Verbunt, F. 1995, in IAU Symp. 165: Compact Stars in 
Binaries, ed. J. van Paradijs, E. P. J. van den Heuvel, \&  E.
Kuulkers (Dordrecht: Kluwer), 333

\bibitem[]{} Wheeler, J. C. 1996, in
Relativistic Astrophysics: A Conference in Honor of Igor Novikov's
60th Birthday", ed. B. Jones \& D. Markovi\'c 
(Cambridge, UK: Cambridge Univ. Press), 211


\bibitem[]{} Wood, J, Horne, K., Berriman, G., Wade, R., O'Donoghue,
D., \& Warner, B. 1986, \mnras, 219, 629

\bibitem[]{} Wood, J, Horne, K., Berriman, G., \& Wade, R. 1989, \apj,
341, 974

\bibitem[]{} Yi, I., Narayan, R., Barret, D., \& McClintock,
J. E. 1996, \aaps, 120C, 187




\end{thebibliography}
\end{document}